\documentclass{article}

\usepackage[preprint]{neurips_2026}
\usepackage[utf8]{inputenc} 
\usepackage[T1]{fontenc}    
\usepackage{hyperref}       
\usepackage{url}            
\usepackage{booktabs}       
\usepackage{amsfonts}       
\usepackage{nicefrac}       
\usepackage{microtype}      
\usepackage{xcolor}         
\usepackage{graphicx}

\usepackage{subcaption}
\usepackage{wrapfig}
\usepackage[most]{tcolorbox}
\usepackage{listings}

\usepackage{tikz}
\usetikzlibrary{positioning, arrows.meta, fit, calc}

\lstdefinestyle{toml}{
  basicstyle=\ttfamily\small,
  backgroundcolor=\color{gray!15},
  frame=single,
  rulecolor=\color{black},
  framesep=4pt,
  xleftmargin=2pt,
  xrightmargin=2pt,
  showstringspaces=false,
  columns=fullflexible,
}

\title{Emulating Heterogeneous Client Execution in Federated Learning}
\author{%
  Arno Geimer \\
  Independent Researcher  \\
  \texttt{research.arge@pm.me} \\
  \And
  Beltran Fiz \\
  University of Luxembourg \\
  \And
  Dimitris Stripelis \\
  Flower Labs
}

\date{\today}

\begin{document}

\maketitle

\begin{abstract}
Federated learning systems are inherently subject to client heterogeneity arising from differences in hardware capabilities. We propose a realistic evaluation framework for hardware-aware federated learning methods based on lightweight emulation of client hardware. Existing evaluation approaches address this either through small-scale real-device deployments or through trace-driven and probabilistic simulations. The former are difficult to scale and reproduce, while the latter suffer from three compounding sources of uncertainty: the choice of distribution family, the choice of execution-time estimates, and the inability to capture workload-dependent behavior. Our framework reproduces heterogeneous client behavior at execution time, capturing compute capacity, memory-constrained feasibility, and their interaction within a single host system. Unlike fixed-trace or throughput-scaling approaches, the proposed method is workload-aware: the same client population can exhibit different runtime and failure behavior depending on the task under study. We evaluate the fidelity of the approach across multiple workloads and show that it preserves relative device performance while accurately reflecting hardware-dependent execution constraints. Compared against direct hardware measurements and AI Benchmark references, the emulation accurately reproduces per-batch feasibility and training performance, preserving both relative device ordering and absolute compute times. We further demonstrate the importance of workload-aware compute times by evaluating different heterogeneity-management methods across multiple workloads. By coupling emulation with real-world-based device sampling, our framework enables realistic, scalable, and reproducible evaluation of federated learning systems under heterogeneous learning conditions, while providing a practical way to generate workload-specific runtime behavior for large and diverse client populations.

\end{abstract}

\section{Introduction}

Federated Learning (FL) is a decentralized machine learning paradigm in which multiple clients collaboratively train a shared model without exchanging raw data~\cite{mcmahan2017communication}. A central challenge in FL is heterogeneity, arising from differences in data distribution, client availability, system architectures, and hardware capabilities~\cite{chen2025advances, kairouz2021advances, wang2021field}.

\definecolor{nvidiagreen}{RGB}{118,185,0}
\definecolor{paramscolor}{RGB}{252,204,204}   
\definecolor{setupcolor}{RGB}{194,232,247}    
\definecolor{workloadcolor}{RGB}{220,223,240} 
\definecolor{tomlcolor}{RGB}{238,238,239}     
\usetikzlibrary{calc}

\newcommand{\annot}[1]{%
  \tikz[baseline=-0.6ex]\node[circle,draw=black,fill=white,text=black,font=\bfseries\small,inner sep=1.5pt,minimum size=12pt]{#1};\,%
}

\begin{figure}[t]
\centering
\begin{tikzpicture}[
    font=\small,
    >=Latex,
    outerbox/.style={
        draw,
        rounded corners=1pt,
        thick,
        minimum width=0.3\textwidth,
        minimum height=5.4cm,
        align=left
    },
    innerbox/.style={
        draw,
        thick,
        minimum width=3.2cm,
        minimum height=0.9cm,
        align=center,
        fill=gray!8
    },
    tomlbox/.style={
        draw, fill=tomlcolor,
        align=left, inner sep=5pt
    },
    paramsbox/.style={
        draw, fill=paramscolor,
        align=center, inner sep=5pt
    },
    lockbox/.style={
        draw, fill=setupcolor,
        align=left, inner sep=5pt, minimum width=3.5cm
    },
    mpsbox/.style={
    draw, fill=white,
    align=center, inner sep=2pt
    },
    workloadbox/.style={
        draw, fill=workloadcolor,
        align=left, inner sep=5pt, minimum width=3.5cm
    },
    annotnode/.style={
        circle, draw=black, fill=white, text=black,
        font=\bfseries\small, inner sep=1.5pt, minimum size=12pt
    },
]

\node[outerbox] (phase1) {};
\node[outerbox, right=5mm of phase1] (phase2) {};
\node[outerbox, right=5mm of phase2] (phase3) {};

\node[anchor=north west, font=\bfseries] at ([xshift=1mm,yshift=-0.5mm]phase1.north west)
    {A. Emulation preparation};
\node[anchor=north west, font=\bfseries] at ([xshift=0mm,yshift=-0.5mm]phase2.north west)
    {B. Workload execution env};
\node[anchor=north west, font=\bfseries] at ([xshift=1mm,yshift=-0.5mm]phase3.north west)
    {C. Output};


\node[tomlbox] (config) at ([yshift=-15.25mm]phase1.north) {
\ttfamily\small
\textbf{[client\_0]}\\[1pt]
\begin{tabular}{@{}l@{\ =\ }l@{}}
gpu    & "RTX 3060"     \\
cpu    & "Ryzen 5 5600" \\
ram\_gb & 16
\end{tabular}
};
\node[lockbox, below=2mm of config] (syslimit) {
    \textbf{Limit system-wide}\\[2pt]
    GPU clock (\texttt{nvidia-smi})\\
    CPU clock (\texttt{cpupower})
};
\node[font=\small\bfseries, above=0.1cm of config] (configlabel) {};
\node[paramsbox, below=2mm of syslimit] (inparams) {
    \textbf{Parameters}\\[1pt]
    global weights
};


\node[mpsbox] (MPS) at ([yshift=-14mm]phase2.north){
\begin{tikzpicture}[baseline=0pt]
\coordinate (mpsgrid) at (0,0);
\node[font=\small] at ($(mpsgrid) + (0, 0.6)$) {\texttt{CUDA MPS}};
\def\cellw{.45cm}
\def\cellh{.15cm}
\node [rectangle, draw, fill=white, minimum width=6*\cellw, minimum height=6*\cellh] at (mpsgrid) {};
\node [rectangle, draw, fill=nvidiagreen, minimum width=3*\cellw, minimum height=6*\cellh] at ($(mpsgrid) + (-1.5*\cellw, 0)$) {};
\node [rectangle, fill=nvidiagreen, draw, minimum width=\cellw, minimum height=3*\cellh] at ($(mpsgrid) + (.5*\cellw, 1.5*\cellh)$) {};
\node [rectangle, fill=gray, draw, minimum width=\cellw, minimum height=3*\cellh] at ($(mpsgrid) + (.5*\cellw, -1.5*\cellh)$) {};
\node [rectangle, draw, fill=gray, minimum width=2*\cellw, minimum height=6*\cellh] at ($(mpsgrid) + (2*\cellw, 0)$) {};
\draw (-0.9, -0.45) -- (-0.9, 0.45);
\draw (-0.45, -0.45) -- (-0.45, 0.45);
\draw (0, -0.45) -- (0, 0.45);
\draw (0.45, -0.45) -- (0.45, 0.45);
\draw (0.9, -0.45) -- (0.9, 0.45);
\draw (-1.35, -0.45) -- (1.35, -0.45);
\draw (-1.35, -0.3) -- (1.35, -0.3);
\draw (-1.35, -0.15) -- (1.35, -0.15);
\draw (-1.35, 0) -- (1.35, 0);
\draw (-1.35, 0.15) -- (1.35, 0.15);
\draw (-1.35, 0.3) -- (1.35, 0.3);
\draw (-1.35, 0.45) -- (1.35, 0.45);
\end{tikzpicture}
};

\node[lockbox, below=0.1cm of MPS] (proclimit) {
    \textbf{Limit process-wide}\\[2pt]
    VRAM (\texttt{torch.cuda})\\
    CPU cores (\texttt{os})
};

\node[workloadbox, below=0.2cm of proclimit, anchor=north] (WORKLOAD) {
    \textbf{Task module}\\[2pt]
    \texttt{load\_data()}\\
    \texttt{train()}
};
\node[lockbox] (reset) at ([yshift=-14mm]phase3.north) {
    \textbf{Reset system-wide}
};
\node[tomlbox, below=2mm of reset] (timings) {
\textbf{[result]}\\[1pt]
\begin{tabular}{@{}l@{\ =\ }l@{}}
preprocess & 1.24\,s \\
training   & 8.43\,s \\
oom        & false
\end{tabular}
};
\node[paramsbox, below=3mm of timings] (outparams) {
    \textbf{Parameters}\\[1pt]
    model update
};

\draw[->, double, thick] (phase1.east) -- (phase2.west);
\draw[->, thick] (inparams.east)
    -- ($(inparams.east)!0.7!(WORKLOAD.west |- inparams.east)$)
    |- (WORKLOAD.west);
\draw[->, thick] (WORKLOAD.east)
    -- ($(WORKLOAD.east)!0.4!(timings.west |- WORKLOAD.east)$)
    -- (outparams.west);
\draw[->, thick] (WORKLOAD.east)
    -- ($(WORKLOAD.east)!0.4!(timings.west |- WORKLOAD.east)$)
    |- (timings.west);

\node[annotnode] at ([xshift=0mm, yshift=6mm]config.east)   {1};
\node[annotnode] at ([xshift=-1mm, yshift=-6mm]syslimit.east) {2};
\node[annotnode] at ([xshift=0mm, yshift=4mm]inparams.east) {3};

\node[annotnode] at ($(phase1.east)!0.5!(phase2.west)+(0.25mm,4mm)$) {4};
\node[annotnode] at ([xshift=0mm, yshift=5mm]MPS.east)       {5};
\node[annotnode] at ([xshift=0mm, yshift=-5mm]proclimit.east) {6};
\node[annotnode] at ([xshift=-1mm, yshift=-6mm]WORKLOAD.east)  {7};
\node[annotnode] at ([xshift=1mm, yshift=-5mm]outparams.east) {3};
\node[annotnode] at ([xshift=0mm, yshift=6mm]timings.east)   {8};

\end{tikzpicture}

\vspace{2mm}
\noindent\footnotesize
\begin{tabular}{@{}p{0.48\linewidth}p{0.48\linewidth}@{}}
\annot{1} Client configuration: CPU, GPU, RAM.   & \annot{5}  \texttt{CUDA MPS} reduces subprocess core usage.  \\[2pt]
\annot{2} Clock limits.  & \annot{6} Process-specific VRAM and core limit.   \\[2pt]
\annot{3} Model input/output as torch tensors.& \annot{7} Run user-defined training. \\[2pt]
\annot{4} \texttt{systemd-run} spawn with \texttt{MemoryMax} (RAM) .  & \annot{8} Return workload execution times and potential errors. \\
\end{tabular}

\caption{Overview of the emulation pipeline. 
\textbf{A:} The host machine is profiled and system-level resource constraints are applied to emulate a target device. Model parameters are checkpointed. \textbf{B:} The workload is executed in an isolated subprocess with additional process-level constraints, and execution time is recorded. \textbf{C:} After execution, constraints are released and timing information and updated parameters are returned to the client.}
\label{fig:emulation_pipeline}

\end{figure}

Existing approaches model hardware heterogeneity either through real-device deployments or through trace-driven and probabilistic simulations~\cite{woisetschlager2024fledge, chang2025flash}. These studies typically focus on edge devices such as mobile phones and single-board computers, while more complex multi-component systems such as desktop or workstation-class machines are considered less frequently. Testbed deployments provide high realism, but they are difficult to scale, expensive to maintain, and often hard to reproduce. Simulation-based approaches are easier to use, but typically rely on abstract performance models that assume workloads differ only in execution time. This gap is consequential as FL workloads increasingly target consumer-grade hardware. On-device fine-tuning of language models, personalization of locally trained agents, and user-adaptive assistants are pushing FL workloads onto desktop and workstation-class machines whose CPU–RAM–GPU interactions differ substantially from the mobile and single-board-computer settings that dominate existing FL benchmarks~\cite{wu2024fedbiot, ji2024emerging}.

This creates a trade-off between fidelity and accessibility. Trace-driven and probabilistic simulations may not cover diverse hardware populations and often ignore workload-dependent execution behavior. Conversely, real-device studies across heterogeneous client populations require substantial hardware and operational resources, which are often unavailable in research settings. We address this gap by introducing a framework-agnostic emulation engine that reproduces heterogeneous client behavior at the level of execution. Rather than aiming for exact architectural reproduction, the proposed approach captures the dominant factors governing workload feasibility and performance behavior with sufficient fidelity for system-level FL evaluation.

Our framework focuses on multi-component devices composed of interacting resources such as CPU, RAM, and GPU. A detailed breakdown of the emulation framework is given in Figure~\ref{fig:emulation_pipeline}. Device behavior is modeled through reproducible, profile-based configurations and system-level resource controls. This enables realistic emulation of heterogeneous FL environments on a single host system without requiring access to diverse physical hardware. More importantly, the induced heterogeneity is workload-aware: The same device population can exhibit different runtime distributions and different feasible execution regions depending on the workload. This is particularly relevant in FL, where scheduling, client participation, and straggler behavior depend not only on how fast clients are, but also on whether a workload can be executed at all. Our proposed framework provides a reproducible and accessible alternative to real-device deployments while capturing system-level effects that are typically abstracted in simulation-based approaches.

\paragraph{Our contributions are as follows.} First, we propose a lightweight, \emph{workload-aware emulation methodology} for modeling heterogeneous client execution in federated learning, enabling realistic evaluation of scheduling strategies and system behavior. Second, we empirically demonstrate that the method \emph{preserves relative device performance} across workloads while capturing hardware-dependent execution constraints, including workload-dependent memory failures that are not reflected in performance-only heterogeneity models. Third, we show that replacing abstract runtime models with workload-specific, execution-level client emulation \emph{can alter the conclusions} drawn about heterogeneity-aware FL methods.

\section{Related Work}

We review related work along three main directions: studies that evaluate federated learning methods on heterogeneous hardware, frameworks that support experimentation under heterogeneous conditions, and device simulation approaches in FL and other domains.

\paragraph{Hardware heterogeneity in Federated Learning.} In FL literature, ``heterogeneous hardware'' refers to clients differing in compute throughput, memory capacity, availability, connection latency, and/or energy behavior. In practice, these differences manifest as per-client variation in local training latency, ability to meet synchronization deadlines, and, in some cases, the fraction or size of a model that can be trained locally. Such heterogeneity has been shown to significantly impact the overall performance and convergence behavior of FL systems~\cite{abdelmoniem2021impact}.

Multiple works evaluate their methods on actual heterogeneous hardware. These evaluations use a variety of hardware configurations, most commonly including single-board computers such as Raspberry Pis, PandaLatte, and Jetson Nanos~\cite{shin2024effective,fan2024behave}. Other works, particularly those focusing on edge-device Federated Learning, additionally include mobile phones in their hardware setups~\cite{geng2025fedex,shin2022fedbalancer,mathur2021device}. An interesting direction is taken in~\cite{bovzivc2024testbed}, which proposes a publicly accessible testbed of more than 40 devices for Federated Learning research. Local simulation of heterogeneous client capacities has been proposed in~\cite{zhang2023fedhc}, although the simulation is limited to GPU capacity fragmentation. A similar approach is taken in~\cite{chai2020tifl, yang2022edgetb, symeonides2023fedbed}, where heterogeneity is induced through grouping of CPU cores. Real-world heterogeneous cross-silo federations have also been implemented by leveraging differences in GPU and CPU compute capabilities~\cite{stripelis2022semi}. An alternative to experimentation on actual heterogeneous hardware is to infer device heterogeneity from real traces. ~\cite{yang2022flash,yang2021characterizing} introduce a large-scale real-world device-state trace and capacity dataset collected from mobile devices, on which~\cite{zhang2024flhetbench} build FLHetBench, a benchmark for device heterogeneity. FLHetBench and similar trace-driven approaches differ from our setting along three axes: they target mobile device populations rather than multi-component desktop-class systems, they replay recorded runtimes rather than execute workloads under resource constraints, and they treat each client as characterized by a fixed performance profile that is independent of the task under study. As a result, they cannot express workload-dependent feasibility or capture how the same client population behaves differently across workloads. Latency is simulated in~\cite{nishio2019client} by uniformly distributing clients around an edge server. More generally, client heterogeneity in mobile devices is often simulated using probabilistic distributions over client resources and response times~\cite{charles2021large, jiang2022pisces}.

\paragraph{Heterogeneity in Frameworks.} Beyond algorithmic contributions and benchmark design, several Federated Learning frameworks have been proposed to support experimental evaluation under realistic system conditions. Existing platforms differ significantly in their ability to model client heterogeneity.
General-purpose frameworks such as \texttt{pfl-research}~\cite{granqvist2024texttt}, Nvidia Flare~\cite{roth2022nvidia}, and Flower~\cite{beutel2020flower} primarily target the implementation and evaluation of learning algorithms, and assume user-defined client behavior, placing the burden of modeling heterogeneity on the experimenter. More recent simulation-oriented systems, including FedScale~\cite{lai2022fedscale}, and FLUTE~\cite{garcia2022flute}, incorporate coarse-grained representations of heterogeneity through device traces and abstract runtime parameters, enabling large-scale experimentation under more realistic condition.
Existing frameworks typically replay traces or assign abstract client runtimes rather than enforce hardware-level behavior. As a result, FedScale and FLUTE can represent aggregate differences in client runtime but do not execute workloads under enforced hardware constraints, and thus do not capture intra-device resource limitations, workload-dependent compute variability, or memory-induced infeasibility, which can all significantly influence training dynamics in heterogeneous environments.

\paragraph{Hardware simulation across domains.} In application testing, device emulation is widely adopted due to the difficulty of accessing a diverse range of hardware platforms. This is particularly evident in mobile systems, where emulators are routinely used to reproduce device-specific conditions on general-purpose computing hardware~\cite{androidsimulator, iossimulator}. In the context of edge computing, testbed architectures commonly rely on virtualization technologies (Docker containers or virtual machines), often combined with an emulated network layer to approximate real-world deployment conditions~\cite{pfandzelter2024lessons}. Within such environments, node characteristics are typically defined through configuration files and enforced by the underlying virtualization engine~\cite{hasenburg2021mockfog}. However, existing approaches primarily focus on constraining CPU and storage resources, with limited support for more complex hardware components.

GPU throttling has predominantly been explored in the context of reducing energy consumption in deep learning workloads~\cite{krzywaniak2022gpu, kakolyris2025throttll}. While these approaches demonstrate that GPU performance can be modulated, they are generally not designed to accurately emulate lower-tier hardware. More precise GPU behavior can be reproduced using device simulators, frequently employed in architecture research and small-scale workload analysis. These simulators capture low-level hardware characteristics, such as kernel execution behavior and cache hierarchies~\cite{bakhoda2009analyzing}. For instance,~\cite{khairy2020accel} introduce a trace- and execution-driven GPU simulator validated across multiple consumer devices. However, such simulators are primarily intended for fine-grained analysis of kernel execution and microarchitectural behavior. As a result, they often incur significantly higher execution overhead compared to native execution on commodity runtimes, making them less suitable for large-scale machine learning workloads~\cite{lew2019analyzing}. This highlights a trade-off between simulation fidelity and execution efficiency, where high-fidelity simulators incur prohibitive overhead, while lightweight approaches lack accuracy. While~\cite{DeepBench} provides a benchmark for training performance evaluation and prediction, it focuses on low-level computational kernels and does not capture the characteristics of modern deep learning workloads or full training pipelines.

Across these directions, existing approaches either rely on real deployments, trace-based abstractions, or high-fidelity simulation, but to the best of our knowledge do not provide a general-purpose mechanism to emulate heterogeneous hardware behavior for system-level FL experimentation.

\section{Method}

We propose an emulation engine for multi-component systems with the goal of improving the realism of heterogeneity evaluation in federated learning. Using profile-based configurations, each client workload can be executed through the emulation engine, covering the full pipeline from data loading and preprocessing to model training.

Combined with a network-level simulation model (see Appendix~\ref{appendix:network} for details), the approach is implemented as a plug-in extension to the Flower framework. It requires minimal additional dependencies and enables straightforward deployment across diverse machine learning workloads.

\subsection{Emulation Engine}

The engine is designed to operate within standard machine learning runtimes, requiring no additional dependencies beyond common Linux and CUDA tooling. The design of the emulation engine is guided by three objectives: (i) lightweight execution without relying on high-fidelity hardware simulators, (ii) compatibility with any Python workloads, and (iii) ease of integration into existing federated learning frameworks. Rather than explicitly modeling hardware architectures, we approximate their observable performance characteristics through controlled resource constraints and execution-time modulation. See Appendix~\ref{appendix:hw_db} for more details on hardware profiles and specific resource constraints.

The emulation engine leverages standard system-level controls and external tooling to approximate lower-end compute capabilities. After profiling the host system, compute resources are scaled relative to the target device configuration to mirror its hardware characteristics. We emulate heterogeneous devices by controlling three aspects of execution: CPU capability, memory availability, and GPU performance. CPU constraints primarily affect data loading and preprocessing throughput, RAM acts as a feasibility constraint on admissible workloads, and GPU constraints affect both training performance and feasibility. In particular, the engine enforces constraints on system memory and GPU memory, processor clock speeds, CPU core availability, and GPU compute capacity and clock speeds. Workloads are executed within a dedicated subprocess, isolating their resource allocation from the main federated learning process and enabling controlled enforcement of device-specific constraints. All experiments disable TF32 and tensor-core paths, ensuring that execution differences across GPUs are attributable to CUDA-core throughput, memory bandwidth, and cache behavior rather than tensor-core presence.

\paragraph{Execution Flow}

In our emulation, each client is associated with a hardware profile, either sampled automatically from a representative hardware database (Appendix~\ref{appendix:hw_db}) or specified manually in a TOML configuration file. This profile defines the target device in terms of GPU, CPU, RAM, and geographic location (Figure~\ref{fig:emulation_pipeline}).

The emulation engine can be used in two complementary ways. It can act as a profiling layer for generating workload-specific runtime behavior under heterogeneous hardware profiles, which is useful for evaluating scheduling, topology design, and other system-level FL methods. Additionally, through integration into the Flower framework, it enables full end-to-end experiments in which hardware heterogeneity interacts with changes in datasets, preprocessing pipelines, model architectures, and communication conditions. From Flower’s perspective, the client behaves identically to a standard client, making the emulation transparent to the framework. In both cases, the framework provides a reproducible and accessible alternative to real-device deployments while capturing system-level effects that are typically abstracted in simulation-based approaches.

Across experiments, the wall-clock runtime overhead introduced by the emulation wrapper is constant per client, amounting to $5.9\pm0.1$\,s for ResNet-18 (47MB) and $8.6\pm0.4$\,s for DistilGPT-2 (328MB), independent of training duration. The difference is primarily attributable to model checkpointing I/O via temporary files. In standalone profiling mode, each distinct hardware profile is emulated once to obtain workload-specific runtime statistics, which can then be reused across arbitrarily many clients sharing that profile. A large population of clients will typically contain a limited number of distinct profiles, so profiling cost is bounded by the number of unique profiles rather than the client count, and can additionally be amortized across experiments sharing the same workload.

\paragraph{Automated Device Sampling}\label{sec:sampling}

In contrast to prior work, which typically models client heterogeneity by sampling abstract performance values from statistical distributions or trace-based datasets, we adopt a device-centric approach based on reproducible hardware profiles. We construct configurations that reflect real-world devices and enforce their characteristics through system-level controls, enabling consistent and repeatable experimentation. To capture realistic population structure, we build a hardware sampler based on a real-world device distribution dataset. This enables realistic reproduction of real-world federation setups. More information can be found in Appendix~\ref{appendix:steam_ds}.

Figure~\ref{fig:throughput_distribution} shows the resulting distribution of GPU performance (measured in FLOPs). We compare the empirical distribution to representative synthetic models used in prior work, including Pareto and half-normal distributions~\cite{li2023plato, nguyen2022federated}. The Pareto distribution does not capture the concentration of devices in the low–mid performance range, whereas the half-normal distribution underestimates the distribution head.

\section{Results}

\subsection{Experimental setup}

All experiments are conducted on a single Linux server running Ubuntu 24.04. The host machine is equipped with an AMD Ryzen 7 1800X (8 physical cores, 16 threads, base clock 3.6\,GHz), $2\times16$\,GB of DDR4 RAM at 2133\,MHz, and an NVIDIA GeForce RTX 4070 SUPER with 12\,GB of GDDR6X VRAM. Python 3.12 is used throughout.

Beyond standard Python packages, the emulation engine relies on the following external dependencies: GPU clock and memory clock locking is performed through \texttt{pynvml}, the Python bindings to the NVIDIA Management Library, which requires a compatible NVIDIA driver. GPU thread-level partitioning uses NVIDIA's Multi-Process Service (MPS) shipped with the NVIDIA driver. CPU frequency scaling requires \texttt{sudo} privileges. RAM limits are enforced by Linux cgroups v2. CPU core pinning and GPU memory fraction limiting are handled through standard OS and PyTorch interfaces.

We evaluate three aspects of our emulation: its ability to capture workload feasibility, its reliability in reproducing training-time behavior, and the effect of CPU emulation on data preprocessing.

\paragraph{Experiments.}

We evaluate the emulation framework on two tasks of different modality and scale.

\textit{Image classification (CIFAR-10).} We train a ResNet-18~\cite{he2016deep} (approximately 11.7\,M parameters) on CIFAR-10~\cite{krizhevsky2009learning}, which consists of 50\,000 training and 10\,000 test images of size $32\times32$ across 10 classes. ResNet-18 is a comparatively lightweight convolutional architecture, making it primarily compute-bound on GPU. In addition, we include ViT-Base/16~\cite{dosovitskiy2020image} (approximately 86\,M parameters) for the emulation validation experiments (Section~\ref{sec:emulation_reliability}). Compared to ResNet-18, ViT-Base/16's compute profile places greater memory and compute demands on the system due to its self-attention and dense feed-forward operations. It therefore acts as a complementary workload for evaluating emulation fidelity across architectures. Training uses the Adam optimizer. Data augmentation includes random cropping (with 4-pixel padding), random rotation ($\pm 15^{\circ}$), and horizontal flipping.

\textit{Instruction tuning (DistilGPT2 + LoRA).} We fine-tune DistilGPT2~\cite{sanh2019distilbert} (approximately 82\,M base parameters) on the Alpaca instruction-tuning dataset~\cite{alpaca} (52\,000 examples) using Low-Rank Adaptation (LoRA)~\cite{hu2022lora} with rank $r=8$, $\alpha=16$, and dropout 0.1, applied to the attention projection layers, with AdamW. This yields approximately 150\,K trainable parameters. Each example is formatted as an instruction--input--response triple and tokenized to a maximum sequence length of 128 int64 tokens. Compared to ResNet-18, the DistilGPT2 workload is more memory-intensive due to the storage of key-value caches and intermediate activations across transformer layers. It also exhibits a different compute profile dominated by dense matrix multiplications in the attention and feed-forward blocks.

\subsection{Workload-specific requirements}\label{sec:comp-mem}

A central question is whether the emulation faithfully reproduces device-level execution constraints beyond differences in compute speed. In modern machine learning workloads, memory capacity frequently constitutes the primary limiting factor, often outweighing raw compute throughput, particularly in transformer-based models and tokenization-heavy pipelines. This factor was not considered in previous FL heterogeneity simulations.

\begin{figure}[h]
  \centering
  \begin{subfigure}[t]{0.45\textwidth}
    \centering
    \includegraphics[height=5cm]{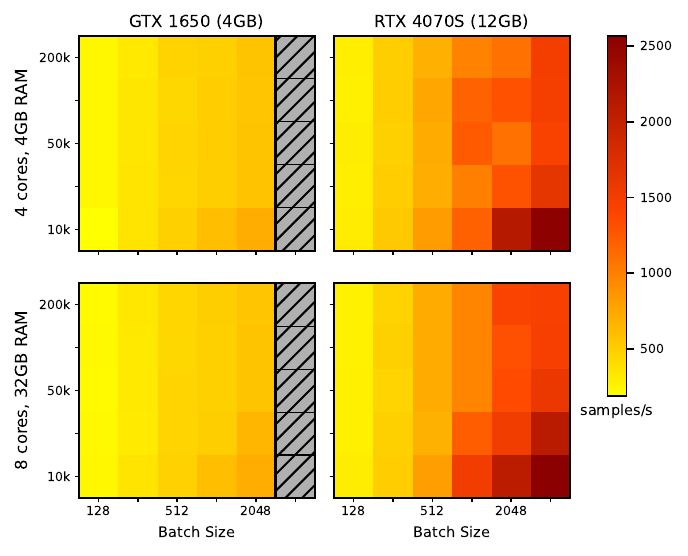}
    \caption{CIFAR-10 (ResNet-18)}
    \label{fig:heatmap_cifar10}
  \end{subfigure}
  \hfill
  \begin{subfigure}[t]{0.5\textwidth}
    \centering
    \includegraphics[height=5cm]{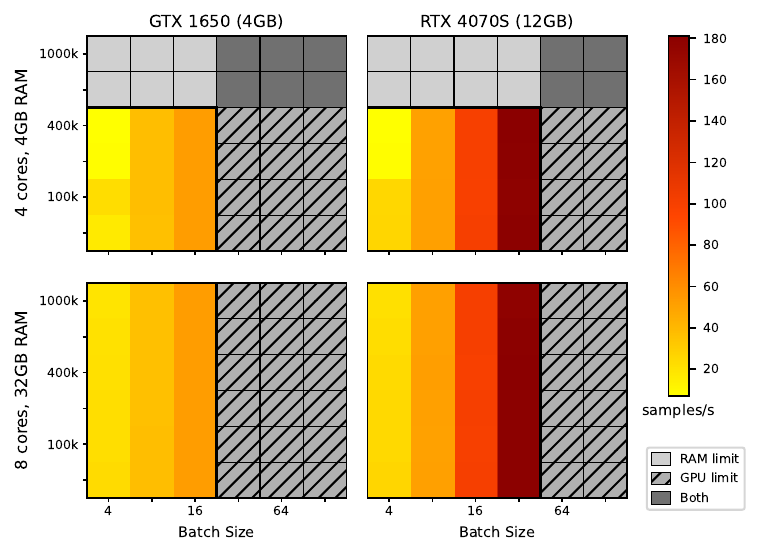}
    \caption{Alpaca (DistilGPT-2)}
    \label{fig:heatmap_llm}
  \end{subfigure}
  \caption{Feasible training regions across batch sizes (x-axis) and dataset sizes (y-axis) under different hardware configurations. Colored regions indicate feasible execution (throughput in samples/s), while shaded regions denote out-of-memory failures due to RAM, GPU, or both. Low-end devices exhibit smaller feasible regions, particularly for LLM workloads.}\label{fig:heatmaps}
\end{figure}
Our emulation explicitly enforces memory limits at both the system (RAM) and GPU (VRAM) levels, enabling the identification of infeasible training configurations. This is particularly relevant in federated learning, where client dropouts and stragglers are often not modeled explicitly or through probabilistic abstractions at best. In contrast, our approach deterministically captures such effects as a function of workload and device characteristics, allowing the precise identification of clients that cannot execute a given task.

Figure~\ref{fig:heatmaps} illustrates memory-related effects for both image classification (Figure~\ref{fig:heatmap_cifar10}) and language model fine-tuning (Figure~\ref{fig:heatmap_llm}). Our emulation captures both RAM- and GPU-induced out-of-memory failures as a function of workload parameters. The correctness of the emulation workload feasibility results is further demonstrated in Table~\ref{tab:hardware-comparison}, with memory errors recorded by real hardware accurately reflected in the emulation. These results demonstrate that hardware heterogeneity does not merely affect training speed, but fundamentally restricts the set of executable configurations. In particular, low-end devices exhibit substantially reduced feasible regions, especially for memory-intensive workloads. This behavior cannot be captured by approaches that model heterogeneity solely through performance scaling.

\subsection{Emulation evaluation}\label{sec:emulation_reliability}

\paragraph{GPU evaluation.}

To assess agreement between emulated and reference values, we report both the Spearman rank correlation coefficient $\rho$, which measures preservation of relative device ordering, and the Pearson correlation coefficient $r$, which quantifies linear agreement and explained variance.

We consider three representative workloads: ResNet-18, ViT-Base, and DistilGPT-2, spanning convolutional and transformer-based vision, as well as language-model training settings. Since the intended use case is the evaluation of hardware-aware federated learning methods, we study both wall-clock reproduction and preservation of relative execution behavior across heterogeneous clients. In FL systems, client ordering, straggler effects, and scheduling outcomes depend on both factors.

Table~\ref{tab:hardware-comparison} compares measured training times on a GTX 1050 (2 GB), a GTX 1080 (8 GB), and the native RTX 4070 SUPER (12 GB) with the corresponding emulation results produced on the RTX 4070 SUPER across workloads. Fidelity decreases at larger batch sizes, likely because newer GPUs, such as the host device, benefit from more efficient memory and caching behavior that cannot be reproduced through resource restriction alone. Still, the agreement is strong overall, especially at smaller batch sizes, indicating that the emulator reliably reproduces training-time behavior across the evaluated workloads.

\begin{table}[h]
\centering
\small
\setlength{\tabcolsep}{4pt}
\caption{Per-batch training time (ms) for three workloads across increasing batch sizes, comparing real hardware against hardware emulation. Values in ms/batch. \textit{OOM} indicates out-of-memory. The emulation tracks real timings with high fidelity: average Spearman $\rho = 0.976$ (Pearson $r = 0.948$).}

\begin{tabular}{l|rrrr|rrrr|rrrr}
\midrule
\textbf{Workload} & \multicolumn{4}{c}{\textbf{ResNet-18}} & \multicolumn{4}{c}{\textbf{ViT-Base}} & \multicolumn{4}{c}{\textbf{DistilGPT-2}} \\
\cmidrule(lr){2-5}\cmidrule(lr){6-9}\cmidrule(lr){10-13}
\textbf{Batch size} & 8 & 16 & 32 & 64 & 8 & 16 & 32 & 64 & 1 & 2 & 4 & 8 \\
\midrule
GTX 1050 & 24.1 & 30.5 & 45.4 & 66.5 & \textit{OOM} & \textit{OOM} & \textit{OOM} & \textit{OOM} & 40.9 & 74.5 & 152.2 & 293.9 \\
GTX 1050 \textit{emul.} & 19.7 & 30.0 & 46.9 & 39.9 & \textit{OOM} & \textit{OOM} & \textit{OOM} & \textit{OOM} & 29.7 & 59.1 & 116.2 & 227.3 \\
\midrule
GTX 1080 & 13.3 & 13.5 & 17.5 & 20.3 & 225.6 & 467.9 & 951.1 & \textit{OOM} & 19.1 & 27.0 & 47.2 & 93.7 \\
GTX 1080 \textit{emul.} & 13.4 & 13.6 & 15.8 & 14.4 & 169.7 & 309.9 & 594.6 & \textit{OOM} & 18.5 & 20.0 & 33.2 & 64.6 \\
\midrule
RTX 4070S & 11.6 & 11.9 & 12.0 & 12.4 & 68.0 & 123.8 & 227.0 & 444.5 & 18.6 & 18.6 & 18.5 & 28.0 \\
RTX 4070S \textit{emul.} & 13.0 & 13.4 & 13.2 & 13.8 & 86.7 & 146.7 & 270.0 & 534.1 & 18.7 & 18.5 & 20.5 & 36.4 \\
\toprule
\end{tabular}
\label{tab:hardware-comparison}
\end{table}

\begin{figure}[h]
  \includegraphics[width=\textwidth]{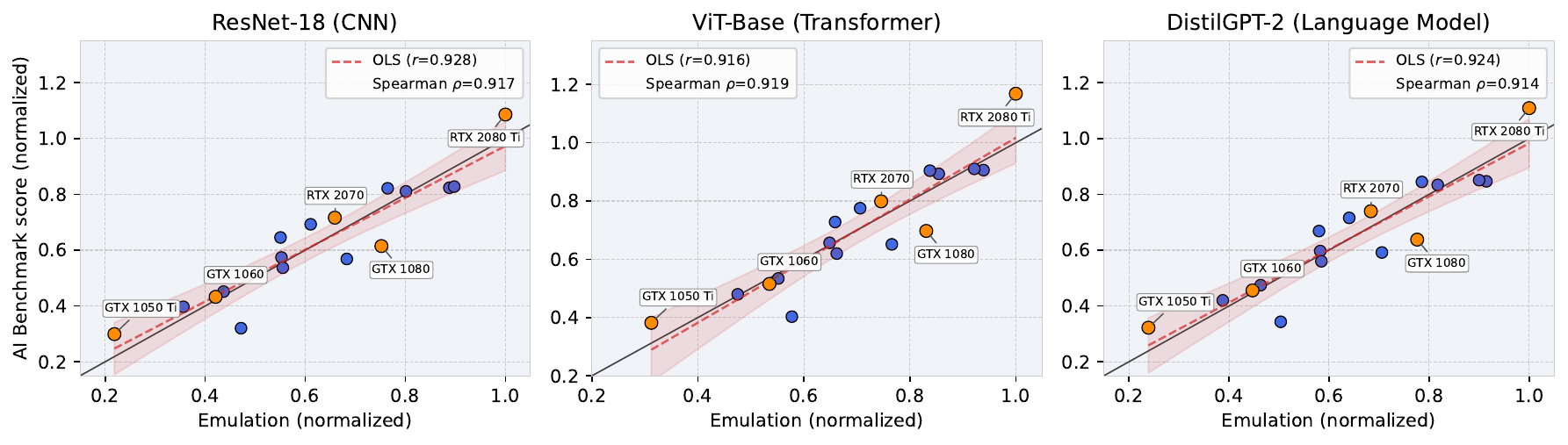}
  \caption{Correlation between expected GPU performance and emulated execution time for 17 devices with 95\% confidence band shaded. Each subplot corresponds to a different model architecture. Points close to the diagonal indicate accurate reproduction of device-specific performance across workloads.}\label{fig:scatt}
\end{figure}

To assess emulation fidelity across a broader range of hardware, we compare emulated GPU training times against AI Benchmark~\cite{aibenchmark}, which we use as a proxy for aggregate device capability. As the benchmark captures end-to-end machine learning performance, it provides a useful reference for relative GPU performance. Figure~\ref{fig:scatt} compares normalized benchmark scores with observed execution performance under emulation across 17 device configurations. Across all workloads, we observe strong agreement ($\rho, r \geq 0.91$), indicating that the emulation reliably preserves relative device behavior. This trend is consistent with the results in Table~\ref{tab:hardware-comparison}, providing further evidence for the fidelity of the proposed emulation framework.
\paragraph{CPU evaluation.}
CPU performance in our emulation is governed by controlled limits on core count and clock frequency enforced at the OS level; we study their effects on the preprocessing stage. 
\begin{wrapfigure}{r}{0.55\textwidth}
  \centering
  \includegraphics[width=\linewidth]{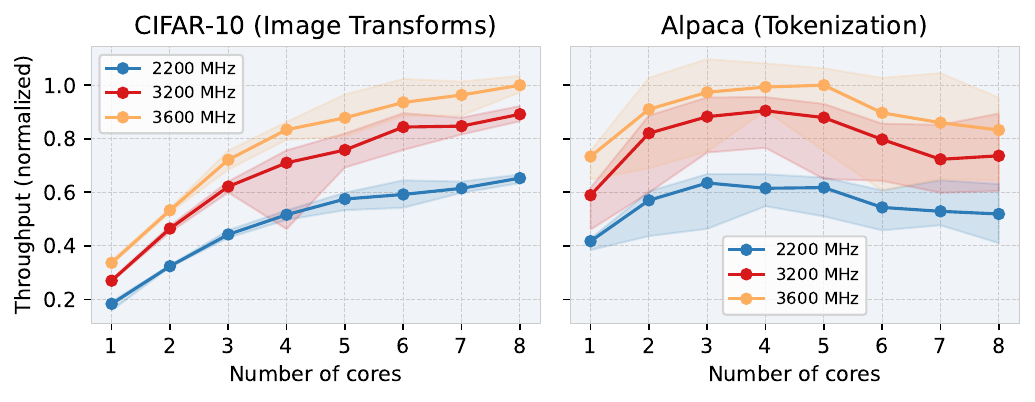}
        \caption{Resource-limited CPU throughput per second of image tensor operations and distilgpt2 tokenization with standard deviations as shaded areas. The tokenization throughput decrease beyond 4 cores may be related to the cache architecture of the host processor.}\label{fig:cpu_bench}
  \vspace{-15pt}
\end{wrapfigure}
Since preprocessing pipelines differ in their CPU and memory demands, we expect workload-dependent behavior. Figure~\ref{fig:cpu_bench} confirms this expectation, showing distinct scaling patterns for image transformations and tokenization across core counts and clock settings. In both workloads, scaling is sublinear, with diminishing returns at higher core counts due to parallelization overheads and data pipeline limitations. This observation matters because preprocessing can bottleneck GPU training by limiting data throughput, making CPU-side emulation an important component of realistic federated learning simulation.

\subsection{The effects of compute time emulation on existing methods}
Having established the fidelity of the proposed emulation, we next showcase why realistic client runtime generation matters for federated learning method evaluation. Specifically, we compare evaluation outcomes obtained with emulated runtimes against those obtained from distribution-based compute models across several heterogeneity-aware FL methods. Our goal is to assess whether conclusions change depending on how client runtimes are modeled. For all experiments, we generate runtimes for 40 different hardware profiles across both ResNet-18 and DistilGPT-2 workloads. Importantly, we observe that the same client population induces different runtime distributions across workloads, indicating that hardware heterogeneity is workload-dependent rather than fixed. This exposes a previously mentioned limitation of trace-driven and distribution-based approaches, which do not capture performance variation induced by task characteristics.

\emph{Throughput-Optimal Topology Design}~\citet{marfoq2020throughput} studies alternative communication topologies for federated learning and reports substantial reductions in overall convergence time relative to the standard star topology under uniform client capabilities. We revisit this setting using our emulated client runtimes as well as distribution-sampled traces. The results are reported in the Appendix (Table~\ref{tab:cross_silo}). While we recover the original findings under uniform client capabilities, the relative speedup of topologies changes once realistic client-side heterogeneity is introduced. We also observe workload-dependent differences in speedup across ResNet-18 and DistilGPT-2. In particular, Ring and $\delta$-MBST achieve higher speedups in larger-scale settings, suggesting improved handling of heterogeneous client runtimes. These results indicate that conclusions drawn from topology evaluations may change once execution-level heterogeneity is taken into account.

\emph{Oort}~\cite{lai2021oort} and \emph{TiFL}~\cite{chai2020tifl} propose alternative client selection strategies under heterogeneous conditions. Oort scores clients using runtime, loss, and data size, whereas TiFL partitions clients into runtime-based tiers and samples across tiers. In the Appendix (Tables~\ref{tab:selection-results-cifar10} and~\ref{tab:selection-results-llm}), we show that different heterogeneity models affect both methods on image classification and language model finetuning. Across all synthetic distributions, timing heterogeneity influences both Oort and TiFL, indicating that it is a fundamental systems cost to manage. However, its effect is workload-dependent: in LLM finetuning, emulated timings allow Oort to outperform the static-timing setting, whereas this does not occur for image classification. Results from CIFAR-10 alone would therefore suggest a different conclusion. This reinforces our main argument that workload-specific compute times are essential for evaluating heterogeneity-aware FL methods, and that realistic timing models can qualitatively alter the conclusions of such evaluations.

In summary, our results show that heterogeneity modeling is not neutral: the choice of runtime model directly affects the observed behavior of FL methods. For realistic approximation of real-world deployments, workload-specific emulated runtimes provide the most informative evaluation signal, as they capture both execution-time variation and workload-dependent behavior that are missed by static or distribution-based timing models.

\section{Discussion}

A central observation of this work is that hardware heterogeneity in federated learning affects both feasibility and runtime, and should therefore not be reduced to execution-time differences alone. Memory constraints can render workloads infeasible on some clients, while workload characteristics can induce different runtime distributions even for the same device population. Both effects are central to FL because they shape client participation, scheduling, and system behavior. Our framework captures heterogeneity as the interaction between workload demands and device constraints, rather than as a single abstract runtime factor. We validate the proposed emulation framework through extensive experiments against real hardware and reference benchmarks and show that it reproduces realistic device behavior on a single host machine. The additional realism introduced through our framework is not merely descriptive. Our case studies show that conclusions drawn under uniform or abstract runtime assumptions can change once realistic client-side heterogeneity is introduced. More broadly, the framework enables controlled evaluation of scheduling strategies, client selection mechanisms, and robustness to stragglers and dropouts.

A limitation of the proposed approach is that by prioritizing scalability and ease of use over high-fidelity hardware simulation, storage performance, cache hierarchies, and low-level kernel behavior are not explicitly modeled. However, the level of fidelity observed is notable and sufficient for realistic federated learning simulation. Finally, the current implementation relies on Linux-based system primitives and CUDA-compatible hardware. These constraints limit portability, but they align well with the primary research setting targeted by this work.

\section{Conclusion}

We presented a lightweight, workload-aware emulation framework for evaluating hardware and communication heterogeneity in federated learning. In contrast to prior approaches that primarily capture heterogeneity through abstract performance scaling or trace-based replay, our method operates at the level of execution, reproducing both differences in computational throughput and hardware-dependent feasibility constraints within a unified framework. Through a profile-based design and system-level resource controls, the proposed approach enables reproducible emulation of heterogeneous client behavior on a single host system. Our evaluation demonstrates that the emulation reliably preserves device performance across diverse workloads while additionally capturing execution constraints, such as memory-induced out-of-memory failures, that fundamentally restrict the set of feasible training configurations. These effects are not captured by traditional trace-driven simulations, but are critical for understanding system behavior in realistic federated settings.

Overall, the results suggest that our approach provides a practical middle ground between high-fidelity simulation and coarse-grained abstractions, enabling scalable, reproducible, and more realistic evaluation of federated learning systems under heterogeneous conditions. We expect this to be particularly useful as FL moves toward on-device fine-tuning and locally trained agents on consumer hardware, where feasibility constraints and workload-specific runtime behavior are central to understanding system performance.

\bibliographystyle{plainnat}
\bibliography{references.bib}

@inproceedings{mcmahan2017communication,
  title={Communication-efficient learning of deep networks from decentralized data},
  author={McMahan, Brendan and Moore, Eider and Ramage, Daniel and Hampson, Seth and y Arcas, Blaise Aguera},
  booktitle={Artificial intelligence and statistics},
  pages={1273--1282},
  year={2017},
  organization={Pmlr}
}

@article{kairouz2021advances,
  title={Advances and open problems in federated learning},
  author={Kairouz, Peter and McMahan, H Brendan},
  journal={Foundations and trends in machine learning},
  volume={14},
  number={1-2},
  pages={1--210},
  year={2021},
  publisher={Emerald Publishing Limited}
}

@article{wang2021field,
  title={A field guide to federated optimization},
  author={Wang, Jianyu and Charles, Zachary and Xu, Zheng and Joshi, Gauri and McMahan, H Brendan and Al-Shedivat, Maruan and Andrew, Galen and Avestimehr, Salman and Daly, Katharine and Data, Deepesh and others},
  journal={arXiv preprint arXiv:2107.06917},
  year={2021}
}

@article{chen2025advances,
  title={Advances in robust federated learning: A survey with heterogeneity considerations},
  author={Chen, Chuan and Liao, Tianchi and Deng, Xiaojun and Wu, Zihou and Huang, Sheng and Zheng, Zibin},
  journal={IEEE Transactions on Big Data},
  volume={11},
  number={3},
  pages={1548--1567},
  year={2025},
  publisher={IEEE}
}

@inproceedings{wu2024fedbiot,
  title={Fedbiot: Llm local fine-tuning in federated learning without full model},
  author={Wu, Feijie and Li, Zitao and Li, Yaliang and Ding, Bolin and Gao, Jing},
  booktitle={Proceedings of the 30th ACM SIGKDD conference on knowledge discovery and data mining},
  pages={3345--3355},
  year={2024}
}

@article{ji2024emerging,
  title={Emerging trends in federated learning: From model fusion to federated x learning},
  author={Ji, Shaoxiong and Tan, Yue and Saravirta, Teemu and Yang, Zhiqin and Liu, Yixin and Vasankari, Lauri and Pan, Shirui and Long, Guodong and Walid, Anwar},
  journal={International Journal of Machine Learning and Cybernetics},
  volume={15},
  number={9},
  pages={3769--3790},
  year={2024},
  publisher={Springer}
}

@article{geng2025fedex,
  title={FedEX: Expediting federated learning over heterogeneous mobile devices by overlapping and participant selection},
  author={Geng, Jiaxiang and Li, Boyu and Qin, Xiaoqi and Li, Yixuan and Li, Liang and Hou, Yanzhao and Pan, Miao},
  journal={IEEE Transactions on Mobile Computing},
  year={2025},
  publisher={IEEE}
}

@inproceedings{shin2024effective,
  title={Effective heterogeneous federated learning via efficient hypernetwork-based weight generation},
  author={Shin, Yujin and Lee, Kichang and Lee, Sungmin and Choi, You Rim and Kim, Hyung-Sin and Ko, JeongGil},
  booktitle={Proceedings of the 22nd ACM Conference on Embedded Networked Sensor Systems},
  pages={112--125},
  year={2024}
}

@article{fan2024behave,
  title={Behave differently when clustering: a semi-asynchronous federated learning approach for iot},
  author={Fan, Boyu and Su, Xiang and Tarkoma, Sasu and Hui, Pan},
  journal={ACM Transactions on Sensor Networks},
  volume={20},
  number={3},
  pages={1--28},
  year={2024},
  publisher={ACM New York, NY}
}

@inproceedings{woisetschlager2024fledge,
  title={Fledge: Benchmarking federated learning applications in edge computing systems},
  author={Woisetschl{\"a}ger, Herbert and Erben, Alexander and Mayer, Ruben and Wang, Shiqiang and Jacobsen, Hans-Arno},
  booktitle={Proceedings of the 25th International Middleware Conference},
  pages={88--102},
  year={2024}
}

@inproceedings{bovzivc2024testbed,
  title={Where is the testbed for my federated learning research?},
  author={Bo{\v{z}}i{\v{c}}, Janez and Faustino, Am{\^a}ndio R and Radovi{\v{c}}, Boris and Canini, Marco and Pejovi{\'c}, Veljko},
  booktitle={2024 IEEE/ACM Symposium on Edge Computing (SEC)},
  pages={249--264},
  year={2024},
  organization={IEEE}
}

@article{mathur2021device,
  title={On-device federated learning with flower},
  author={Mathur, Akhil and Beutel, Daniel J and de Gusmao, Pedro Porto Buarque and Fernandez-Marques, Javier and Topal, Taner and Qiu, Xinchi and Parcollet, Titouan and Gao, Yan and Lane, Nicholas D},
  booktitle={Proceedings of the 3rd MLSys Conference},
  year={2021} 
}

@inproceedings{shin2022fedbalancer,
  title={Fedbalancer: Data and pace control for efficient federated learning on heterogeneous clients},
  author={Shin, Jaemin and Li, Yuanchun and Liu, Yunxin and Lee, Sung-Ju},
  booktitle={Proceedings of the 20th Annual International Conference on Mobile Systems, Applications and Services},
  pages={436--449},
  year={2022}
}

@inproceedings{zhang2024flhetbench,
  title={Flhetbench: Benchmarking device and state heterogeneity in federated learning},
  author={Zhang, Junyuan and Zeng, Shuang and Zhang, Miao and Wang, Runxi and Wang, Feifei and Zhou, Yuyin and Liang, Paul Pu and Qu, Liangqiong},
  booktitle={Proceedings of the IEEE/CVF Conference on Computer Vision and Pattern Recognition},
  pages={12098--12108},
  year={2024}
}

@article{stripelis2022semi,
  title={Semi-synchronous federated learning for energy-efficient training and accelerated convergence in cross-silo settings},
  author={Stripelis, Dimitris and Thompson, Paul M and Ambite, Jos{\'e} Luis},
  journal={ACM Transactions on Intelligent Systems and Technology (TIST)},
  volume={13},
  number={5},
  pages={1--29},
  year={2022},
  publisher={ACM New York, NY}
}

@article{zhang2023fedhc,
  title={FedHC: A Scalable Federated Learning Framework for Heterogeneous and Resource-Constrained Clients},
  author={Zhang, Min and Yu, Fuxun and Yu, Yongbo and Zhang, Minjia and Li, Ang and Chen, Xiang},
  journal={arXiv preprint arXiv:2305.15668},
  year={2023}
}

@inproceedings{chai2020tifl,
  title={Tifl: A tier-based federated learning system},
  author={Chai, Zheng and Ali, Ahsan and Zawad, Syed and Truex, Stacey and Anwar, Ali and Baracaldo, Nathalie and Zhou, Yi and Ludwig, Heiko and Yan, Feng and Cheng, Yue},
  booktitle={Proceedings of the 29th international symposium on high-performance parallel and distributed computing},
  pages={125--136},
  year={2020}
}

@article{yang2022edgetb,
  title={EdgeTB: A hybrid testbed for distributed machine learning at the edge with high fidelity},
  author={Yang, Lei and Wen, Fulin and Cao, Jiannong and Wang, Zhenyu},
  journal={IEEE Transactions on Parallel and Distributed Systems},
  volume={33},
  number={10},
  pages={2540--2553},
  year={2022},
  publisher={IEEE}
}

@inproceedings{symeonides2023fedbed,
  title={Fedbed: Benchmarking federated learning over virtualized edge testbeds},
  author={Symeonides, Moysis and Nikolaidis, Fotis and Trihinas, Demetris and Pallis, George and Dikaiakos, Marios D and Bilas, Angelos},
  booktitle={Proceedings of the IEEE/ACM 16th International Conference on Utility and Cloud Computing},
  pages={1--10},
  year={2023}
}

@inproceedings{yang2021characterizing,
  title={Characterizing impacts of heterogeneity in federated learning upon large-scale smartphone data},
  author={Yang, Chengxu and Wang, Qipeng and Xu, Mengwei and Chen, Zhenpeng and Bian, Kaigui and Liu, Yunxin and Liu, Xuanzhe},
  booktitle={Proceedings of the Web Conference 2021},
  pages={935--946},
  year={2021}
}

@article{yang2022flash,
  title={FLASH: Heterogeneity-aware federated learning at scale},
  author={Yang, Chengxu and Xu, Mengwei and Wang, Qipeng and Chen, Zhenpeng and Huang, Kang and Ma, Yun and Bian, Kaigui and Huang, Gang and Liu, Yunxin and Jin, Xin and others},
  journal={IEEE Transactions on Mobile Computing},
  volume={23},
  number={1},
  pages={483--500},
  year={2022},
  publisher={IEEE}
}

@inproceedings{chang2025flash,
  title={Flash: Federated learning across simultaneous heterogeneities},
  author={Chang, Xiangyu and Ahmed, Sk Miraj and Krishnamurthy, Srikanth V and Guler, Basak and Swami, Ananthram and Oymak, Samet and Roy-Chowdhury, Amit K},
  booktitle={MILCOM 2025-2025 IEEE Military Communications Conference (MILCOM)},
  pages={826--831},
  year={2025},
  organization={IEEE}
}

@inproceedings{nishio2019client,
  title={Client selection for federated learning with heterogeneous resources in mobile edge},
  author={Nishio, Takayuki and Yonetani, Ryo},
  booktitle={ICC 2019-2019 IEEE international conference on communications (ICC)},
  pages={1--7},
  year={2019},
  organization={IEEE}
}

@inproceedings{nguyen2022federated,
  title={Federated learning with buffered asynchronous aggregation},
  author={Nguyen, John and Malik, Kshitiz and Zhan, Hongyuan and Yousefpour, Ashkan and Rabbat, Mike and Malek, Mani and Huba, Dzmitry},
  booktitle={International conference on artificial intelligence and statistics},
  pages={3581--3607},
  year={2022},
  organization={PMLR}
}

@article{charles2021large,
  title={On large-cohort training for federated learning},
  author={Charles, Zachary and Garrett, Zachary and Huo, Zhouyuan and Shmulyian, Sergei and Smith, Virginia},
  journal={Advances in neural information processing systems},
  volume={34},
  pages={20461--20475},
  year={2021}
}

@inproceedings{jiang2022pisces,
  title={Pisces: Efficient federated learning via guided asynchronous training},
  author={Jiang, Zhifeng and Wang, Wei and Li, Baochun and Li, Bo},
  booktitle={Proceedings of the 13th Symposium on Cloud Computing},
  pages={370--385},
  year={2022}
}

@inproceedings{li2023plato,
  title={Plato: An open-source research framework for production federated learning},
  author={Li, Baochun and Su, Ningxin and Ying, Chen and Wang, Fei},
  booktitle={Proceedings of the ACM Turing Award Celebration Conference-China 2023},
  pages={1--2},
  year={2023}
}

@article{roth2022nvidia,
  title={Nvidia flare: Federated learning from simulation to real-world},
  author={Roth, Holger R and Cheng, Yan and Wen, Yuhong and Yang, Isaac and Xu, Ziyue and Hsieh, Yuan-Ting and Kersten, Kristopher and Harouni, Ahmed and Zhao, Can and Lu, Kevin and others},
  journal={arXiv preprint arXiv:2210.13291},
  year={2022}
}

@article{beutel2020flower,
  title={Flower: A friendly federated learning research framework},
  author={Beutel, Daniel J and Topal, Taner and Mathur, Akhil and Qiu, Xinchi and Fernandez-Marques, Javier and Gao, Yan and Sani, Lorenzo and Li, Kwing Hei and Parcollet, Titouan and de Gusm{\~a}go, Pedro Porto Buarque and others},
  journal={arXiv preprint arXiv:2007.14390},
  year={2020}
}

@inproceedings{lai2022fedscale,
  title={Fedscale: Benchmarking model and system performance of federated learning at scale},
  author={Lai, Fan and Dai, Yinwei and Singapuram, Sanjay and Liu, Jiachen and Zhu, Xiangfeng and Madhyastha, Harsha and Chowdhury, Mosharaf},
  booktitle={International conference on machine learning},
  pages={11814--11827},
  year={2022},
  organization={PMLR}
}

@article{garcia2022flute,
  title={Flute: A scalable, extensible framework for high-performance federated learning simulations},
  author={Garcia, Mirian Hipolito and Manoel, Andre and Diaz, Daniel Madrigal and Mireshghallah, Fatemehsadat and Sim, Robert and Dimitriadis, Dimitrios},
  journal={arXiv preprint arXiv:2203.13789},
  year={2022}
}

@article{granqvist2024texttt,
  title={pfl-research: Simulation framework for accelerating research in Private Federated Learning},
  author={Granqvist, Filip and Song, Congzheng and Cahill, {\'A}ine and van Dalen, Rogier and Pelikan, Martin and Chan, Yi Sheng and Feng, Xiaojun and Krishnaswami, Natarajan and Jina, Vojta and Chitnis, Mona},
  journal={Advances in Neural Information Processing Systems},
  volume={37},
  pages={43403--43434},
  year={2024}
}

@misc{androidsimulator,
  title = {Run apps on the Android Emulator
},
  key = {Android},
  howpublished = {\url{https://developer.android.com/studio/run/emulator}},
  note = {Accessed: 2026-03-30}
}

@misc{iossimulator,
  title = {Running your app in Simulator or on a device
},
  key = {iOS},
  howpublished = {\url{https://developer.apple.com/documentation/xcode/running-your-app-in-simulator-or-on-a-device}},
  note = {Accessed: 2026-03-30}
}

@misc{deepbench,
  title = {DeepBench: A Benchmark for Deep Learning Workloads
},
  key = {DeepBench},
  howpublished = {\url{https://github.com/baidu-research/DeepBench}},
  note = {Accessed: 2026-03-30}
}

@misc{steamsurvey,
  title = {Steam Hardware \& Software Survey},
  key = {Valve, 2026},
  howpublished = {\url{https://store.steampowered.com/hwsurvey}},
  note = {Accessed: 2026-03-30},
}

@misc{aibenchmark,
  title = {AI-Benchmark Deeplearning Desktop},
  key = {Ignatov, 2026},
  howpublished = {\url{https://ai-benchmark.com/ranking_deeplearning_detailed.html}},
  note = {Accessed: 2026-04-30},
}

@article{pfandzelter2024lessons,
  title={Lessons Learned from Building Edge Software System Testbeds},
  author={Pfandzelter, Tobias and Bermbach, David},
  journal={arXiv preprint arXiv:2403.16869},
  year={2024}
}

@article{hasenburg2021mockfog,
  title={MockFog 2.0: Automated execution of fog application experiments in the cloud},
  author={Hasenburg, Jonathan and Grambow, Martin and Bermbach, David},
  journal={IEEE Transactions on Cloud Computing},
  volume={11},
  number={1},
  pages={58--70},
  year={2021},
  publisher={IEEE}
}

@inproceedings{krzywaniak2022gpu,
  title={GPU power capping for energy-performance trade-offs in training of deep convolutional neural networks for image recognition},
  author={Krzywaniak, Adam and Czarnul, Pawel and Proficz, Jerzy},
  booktitle={International conference on computational science},
  pages={667--681},
  year={2022},
  organization={Springer}
}

@inproceedings{kakolyris2025throttll,
  title={throttll'em: Predictive gpu throttling for energy efficient llm inference serving},
  author={Kakolyris, Andreas Kosmas and Masouros, Dimosthenis and Vavaroutsos, Petros and Xydis, Sotirios and Soudris, Dimitrios},
  booktitle={2025 IEEE International Symposium on High Performance Computer Architecture (HPCA)},
  pages={1363--1378},
  year={2025},
  organization={IEEE}
}

@inproceedings{bakhoda2009analyzing,
  title={Analyzing CUDA workloads using a detailed GPU simulator},
  author={Bakhoda, Ali and Yuan, George L and Fung, Wilson WL and Wong, Henry and Aamodt, Tor M},
  booktitle={2009 IEEE international symposium on performance analysis of systems and software},
  pages={163--174},
  year={2009},
  organization={IEEE}
}

@inproceedings{khairy2020accel,
  title={Accel-sim: An extensible simulation framework for validated gpu modeling},
  author={Khairy, Mahmoud and Shen, Zhesheng and Aamodt, Tor M and Rogers, Timothy G},
  booktitle={2020 ACM/IEEE 47th Annual International Symposium on Computer Architecture (ISCA)},
  pages={473--486},
  year={2020},
  organization={IEEE}
}

@inproceedings{lew2019analyzing,
  title={Analyzing machine learning workloads using a detailed GPU simulator},
  author={Lew, Jonathan and Shah, Deval A and Pati, Suchita and Cattell, Shaylin and Zhang, Mengchi and Sandhupatla, Amruth and Ng, Christopher and Goli, Negar and Sinclair, Matthew D and Rogers, Timothy G and others},
  booktitle={2019 IEEE international symposium on performance analysis of systems and software (ISPASS)},
  pages={151--152},
  year={2019},
  organization={IEEE}
}

@article{marfoq2020throughput,
  title={Throughput-optimal topology design for cross-silo federated learning},
  author={Marfoq, Othmane and Xu, Chuan and Neglia, Giovanni and Vidal, Richard},
  journal={Advances in Neural Information Processing Systems},
  volume={33},
  pages={19478--19487},
  year={2020}
}

@inproceedings{he2016deep,
  title={Deep residual learning for image recognition},
  author={He, Kaiming and Zhang, Xiangyu and Ren, Shaoqing and Sun, Jian},
  booktitle={Proceedings of the IEEE conference on computer vision and pattern recognition},
  pages={770--778},
  year={2016}
}

@article{krizhevsky2009learning,
  title={Learning multiple layers of features from tiny images},
  author={Krizhevsky, Alex and Hinton, Geoffrey and others},
  year={2009},
  publisher={Toronto, ON, Canada}
}

@article{sanh2019distilbert,
  title={DistilBERT, a distilled version of BERT: smaller, faster, cheaper and lighter},
  author={Sanh, Victor and Debut, Lysandre and Chaumond, Julien and Wolf, Thomas},
  journal={arXiv preprint arXiv:1910.01108},
  year={2019}
}

@misc{alpaca,
  author = {Rohan Taori and Ishaan Gulrajani and Tianyi Zhang and Yann Dubois and Xuechen Li and Carlos Guestrin and Percy Liang and Tatsunori B. Hashimoto },
  title = {Stanford Alpaca: An Instruction-following LLaMA model},
  year = {2023},
  publisher = {GitHub},
  journal = {GitHub repository},
  howpublished = {\url{https://github.com/tatsu-lab/stanford_alpaca}},
}

@article{hu2022lora,
  title={Lora: Low-rank adaptation of large language models.},
  author={Hu, Edward J and Shen, Yelong and Wallis, Phillip and Allen-Zhu, Zeyuan and Li, Yuanzhi and Wang, Shean and Wang, Liang and Chen, Weizhu and others},
  journal={Iclr},
  volume={1},
  number={2},
  pages={3},
  year={2022}
}

@inproceedings{
dosovitskiy2020image,
title={An Image is Worth 16x16 Words: Transformers for Image Recognition at Scale},
author={Alexey Dosovitskiy and Lucas Beyer and Alexander Kolesnikov and Dirk Weissenborn and Xiaohua Zhai and Thomas Unterthiner and Mostafa Dehghani and Matthias Minderer and Georg Heigold and Sylvain Gelly and Jakob Uszkoreit and Neil Houlsby},
booktitle={International Conference on Learning Representations},
year={2021},
url={https://openreview.net/forum?id=YicbFdNTTy}
}

@inproceedings{lai2021oort,
  title={Oort: Efficient federated learning via guided participant selection},
  author={Lai, Fan and Zhu, Xiangfeng and Madhyastha, Harsha V and Chowdhury, Mosharaf},
  booktitle={15th $\{$USENIX$\}$ Symposium on Operating Systems Design and Implementation ($\{$OSDI$\}$ 21)},
  pages={19--35},
  year={2021}
}

@article{abdelmoniem2021impact,
  title={On the impact of device and behavioral heterogeneity in federated learning},
  author={Abdelmoniem, Ahmed M and Ho, Chen-Yu and Papageorgiou, Pantelis and Bilal, Muhammad and Canini, Marco},
  journal={arXiv preprint arXiv:2102.07500},
  year={2021}
}

\newpage

\appendix
\section{Network simulation}\label{appendix:network}

\subsection{Network Topology and Communication Modeling}

To approximate realistic global deployment scenarios, we construct a geographically distributed client topology that includes representative countries from each continent. To better reflect real-world usage patterns, we place higher emphasis on regions with strong AI adoption, such as North America, Europe, and parts of Asia (e.g., China and India), resulting in a fully connected, directed networking graph.

For each location, we derive network characteristics from publicly available, up-to-date sources, including median upload and download bandwidth as well as inter-country latency estimates, to approximate communication behavior during training. Model transfer times are computed using the serialized model size in bytes, allowing for a consistent and framework-agnostic estimation of upload and download durations. The modular design of our approach allows additional effects, such as packet loss, to be incorporated by the end user if desired. While the resulting network model is less detailed than other high-fidelity federation simulation environments (~\cite{marfoq2020throughput}), it provides a lightweight and easily extensible approximation of communication heterogeneity. New locations and connectivity patterns can be incorporated with minimal effort, enabling flexible adaptation to different deployment scenarios.

The compute emulation engine operates in conjunction with the network model, which captures communication heterogeneity before and after local training. Specifically, network effects are applied prior to client execution to model model distribution latency, and after execution to approximate upload delays during aggregation. This separation enables independent control of compute and communication characteristics while preserving their interaction within the federated learning workflow.

\subsection{Visualizing a federation round -- from start to finish}

\begin{figure}[h]
  \includegraphics[width=\textwidth]{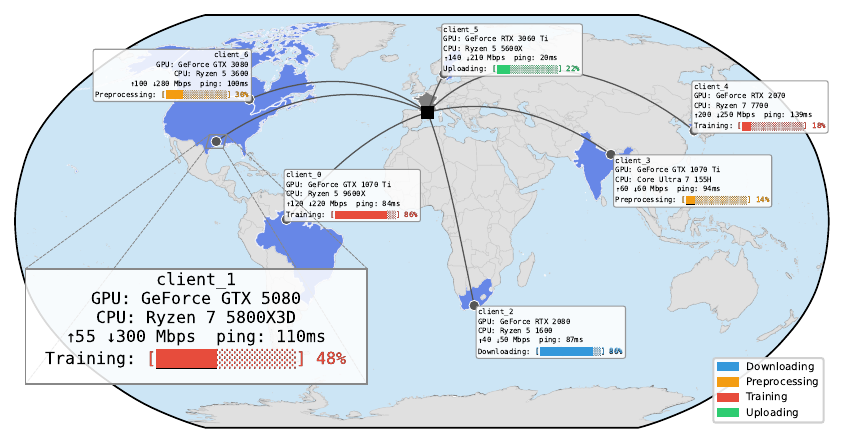}
  \caption{Simulated federated learning round under heterogeneous conditions. Clients distributed across regions exhibit diverse hardware configurations and are at different stages of the training pipeline at a given point in time.}\label{fig:round_map}
\end{figure}

Figure~\ref{fig:round_map} illustrates the full execution pipeline of a federated learning round under our emulation framework. The round is initiated by the central server, which distributes the aggregated global model to participating clients (\textcolor{blue}{Downloading}), with transfer times determined by client-specific bandwidth and latency.

Upon receipt, clients preprocess their local datasets (\textcolor{orange}{Preprocessing}), with performance governed by CPU and RAM constraints, and subsequently perform local training (\textcolor{red}{Training}) on their GPU. After completing local updates, clients transmit their model updates back to the server (\textcolor{green}{Uploading}), again subject to network conditions.

The server aggregates the received updates once all clients have responded—or, alternatively, upon receiving failure notifications—and proceeds to initiate the next round. This visualization highlights the interplay between computation and communication heterogeneity throughout the federated learning process.

\section{Device simulation}

\subsection{Device sampler}\label{appendix:steam_ds}

To generate realistic hardware combinations, we leverage data from the Steam Hardware Survey~\cite{steamsurvey}, a large-scale global measurement of consumer hardware. Although Valve does not release the raw survey data, Steam has over 100 million monthly active users, making the survey a reasonable proxy for consumer hardware prevalence. We sample device configurations according to observed hardware prevalence from both GPU, CPU and RAM distributions. Although the survey may be biased toward gaming-oriented users, its large coverage makes it a reasonable approximation of consumer hardware prevalence.

\begin{figure}[h]
  \includegraphics[width=\textwidth]{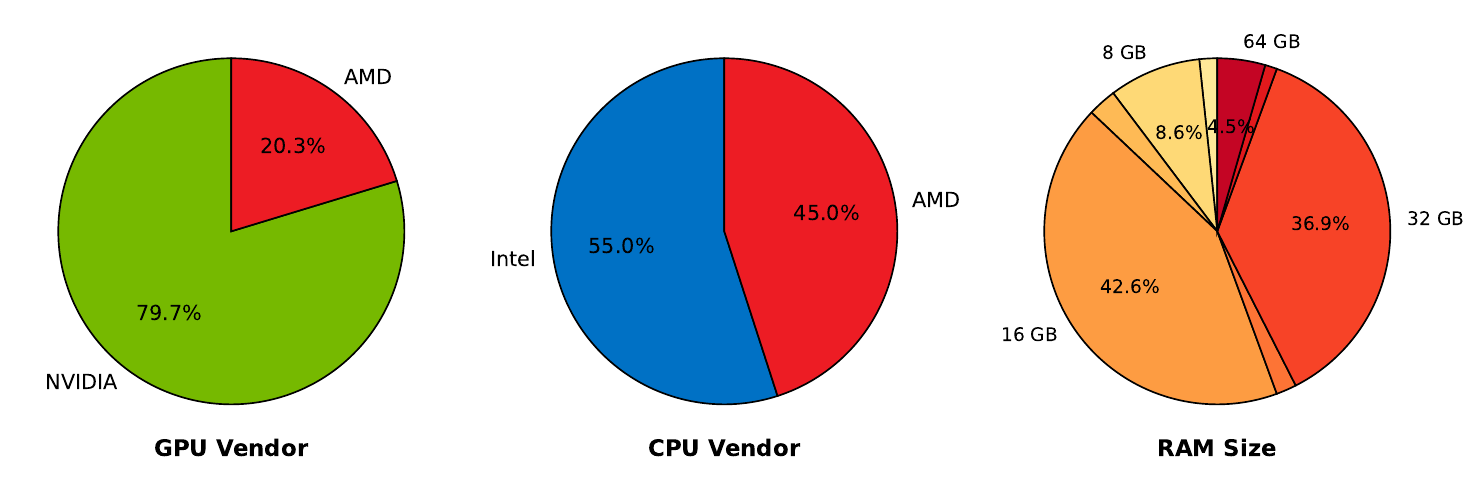}
  \caption{Distribution of GPU vendors, CPU vendors, and RAM capacities among consumer hardware, based on the Steam Hardware Survey (October 2025).}\label{appendix:steam_survey}
\end{figure}

Client profiles are sampled from the hardware database according to device popularity in the survey. An example client profile is shown in Figure~\ref{fig:emulation_pipeline}.

\subsection{Hardware database}\label{appendix:hw_db}

\begin{wrapfigure}{r}{0.4\textwidth}
  \vspace{-15pt}
  \centering
    \includegraphics[width = \linewidth]{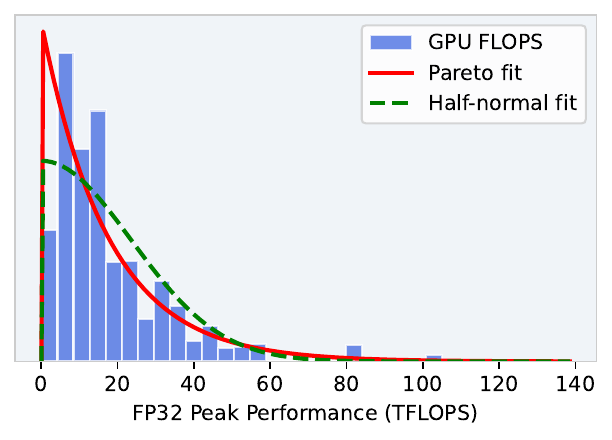}
    \caption{Realistically sampled empirical GPU performance (FLOPs) vs. synthetic heterogeneity models. Both synthetic models fail to capture the low–mid concentration; Pareto overestimates the head of the distribution, while the half-normal fit underestimates it.}\label{fig:throughput_distribution}
 \vspace{-20pt}
\end{wrapfigure}

Device names in client profiles are resolved against a hardware database stored as TOML files. The database contains per-device specifications for both GPUs and CPUs, as illustrated in Figures~\ref{fig:gpu_entry} and~\ref{fig:cpu_entry}.

Each field in the hardware database corresponds to a specific resource constraint enforced by the emulation engine. For GPUs, \texttt{clock\_speed} and \texttt{memory\_speed} are used to lock the GPU core and memory clocks via \texttt{nvidia-smi}; \texttt{cuda\_cores} determines the CUDA MPS active thread percentage, which limits the fraction of GPU compute units available to the worker process; and \texttt{memory\_gb} is enforced through \texttt{torch.cuda.set\_per\_process\_memory\_fraction}, which caps the GPU memory visible to the training workload. For CPUs, the base clock extracted from \texttt{core\_clock} is applied as a system-wide frequency limit via \texttt{sysfs}, while the physical core count from \texttt{cores} is enforced through \texttt{os.sched\_setaffinity}, which pins the worker process to the corresponding number of cores. System RAM (\texttt{ram\_gb} from the client profile) is restricted using Linux cgroups~v2 via \texttt{systemd-run}. The \texttt{shss} field stores the Steam Hardware Survey share used for population-level sampling (Section~\ref{sec:sampling}).

\begin{figure}[h]
\centering
\begin{minipage}[t]{0.47\textwidth}
\centering
\begin{lstlisting}[style=toml]
[[gpus]]
name        = "GeForce GTX 1070"
memory_gb   = 8.0
memory_type = "GDDR5"
memory_bw   = 256.0
clock_speed = 1683.0
memory_speed = 2002.0
cuda_cores  = 1920
shss        = 1.129
\end{lstlisting}
\caption{Example GPU entry in the hardware database.}\label{fig:gpu_entry}
\end{minipage}%
\hfill
\begin{minipage}[t]{0.47\textwidth}
\centering
\begin{lstlisting}[style=toml]
[[cpus]]
name       = "Ryzen 5 3600"
cores      = "6 / 12"
core_clock = "3.6 to 4.2 GHz"
shss       = 0.0163
\end{lstlisting}
\caption{Example CPU entry in the hardware database.}\label{fig:cpu_entry}
\end{minipage}
\end{figure}

\subsection{The effects of compute time emulation on existing methods.}

\begin{wrapfigure}{l}{0.4\textwidth}
  \centering
  \includegraphics[width=\linewidth]{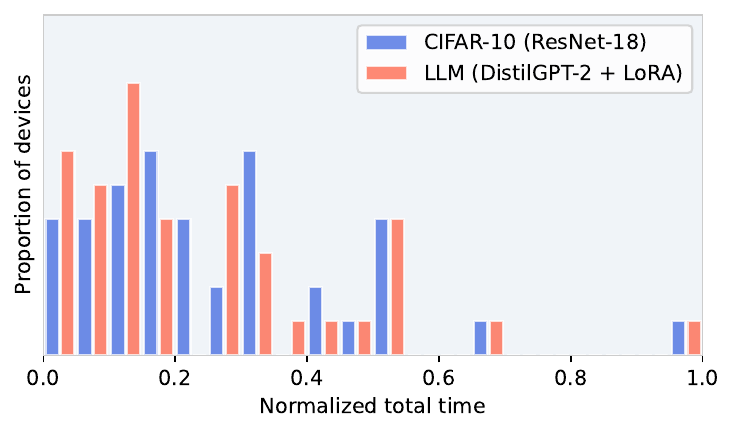}
        \caption{Distribution of normalized wall-clock training time across 40 different clients for ResNet-18 and distilgpt2.}\label{fig:topo_performance_distribution}
\end{wrapfigure}

To study the effects of realistic device compute times, we emulate 40 different client profiles sampled with our device sampler. Figure~\ref{fig:topo_performance_distribution} shows the resulting compute-time distributions. Although the differences between the two workloads are modest, both workloads are lightweight enough to run on all device configurations and still induce distinct runtime behavior. For all experiments, the four timing models share a common mean: the static model assigns every client the average emulated compute time, while the Pareto and half-normal distributions are parameterized so that their mean matches this value. This ensures that differences across timing models reflect the shape of the runtime distribution rather than differences in absolute scale.

\paragraph{Throughput-Optimal Topology Design}. The authors consider three real-world networks and two synthetic ones, and evaluate communication topologies over these networks: the standard \emph{star} topology, consisting of a central server and communicating clients, as well as the decentralized \emph{ring} and $\delta$-MBST topologies, which are designed to reduce computation-induced slowdown.

We reproduce their experimental setup, with one key modification: instead of assuming static training times across nodes, we assign heterogeneous device profiles to clients. Using our hardware sampler, we generate 40 client configurations, emulate training times for both CIFAR-10 and DistilGPT-2, and normalize the resulting timings to match the scale of the original study, namely iNaturalist.

\begin{table}[h]
\centering
\caption{Cycle time (ms) per round under different compute time models. Speedup over Star topology in parentheses. CIFAR-10 and LLM compute times emulated.}
\label{tab:cross_silo}
\begin{tabular}{llrrrrr}
\midrule
Network & Topology & Static & Pareto & Half-Normal & CIFAR-10 & LLM \\
\midrule
  gaia & Star & 391 & 486 & 395 & 436 & 442 \\
   & Ring & 118 (3.3$\times$) & 118 (4.1$\times$) & 118 (3.3$\times$) & 126 (3.4$\times$) & 127 (3.5$\times$) \\
   & $\delta$-MBST & 138 (2.8$\times$) & 140 (3.5$\times$) & 139 (2.8$\times$) & 162 (2.7$\times$) & 165 (2.7$\times$) \\
\midrule
  aws\_na& Star & 189 & 286 & 203 & 248 & 256 \\
   & Ring & 81 (2.3$\times$) & 81 (3.5$\times$) & 81 (2.5$\times$) & 82 (3.0$\times$) & 82 (3.1$\times$) \\
   & $\delta$-MBST & 90 (2.1$\times$) & 120 (2.4$\times$) & 98 (2.1$\times$) & 101 (2.5$\times$) & 104 (2.5$\times$) \\
\midrule
  geant & Star & 622 & 734 & 641 & 663 & 666 \\
   & Ring & 109 (5.7$\times$) & 109 (6.7$\times$) & 109 (5.9$\times$) & 106 (6.2$\times$) & 109 (6.1$\times$) \\
   & $\delta$-MBST & 101 (6.2$\times$) & 119 (6.2$\times$) & 110 (5.9$\times$) & 110 (6.0$\times$) & 113 (5.9$\times$) \\
\midrule
  exodus & Star & 840 & 1167 & 861 & 883 & 886 \\
   & Ring & 104 (8.1$\times$) & 104 (11.2$\times$) & 104 (8.3$\times$) & 104 (8.5$\times$) & 104 (8.5$\times$) \\
   & $\delta$-MBST & 145 (5.8$\times$) & 241 (4.8$\times$) & 142 (6.1$\times$) & 141 (6.3$\times$) & 142 (6.2$\times$) \\
\midrule
  ebone & Star & 690 & 1075 & 712 & 750 & 757 \\
   & Ring & 95 (7.2$\times$) & 95 (11.3$\times$) & 95 (7.5$\times$) & 95 (7.9$\times$) & 96 (7.9$\times$) \\
   & $\delta$-MBST & 122 (5.7$\times$) & 244 (4.4$\times$) & 118 (6.0$\times$) & 117 (6.4$\times$) & 130 (5.8$\times$) \\
\bottomrule
\end{tabular}
\end{table}

\paragraph{Oort and TiFL}. 
As before, we use our 40 emulated client profiles to revisit the conclusions of Oort and TiFL under realistic hardware heterogeneity. Oort ranks clients using a utility score that includes compute time, local loss, and dataset size, meaning that clients with lower compute times may be selected more often. In the original work, these compute times are assigned from traces independently of the workload. TiFL instead groups clients into compute-time tiers and samples clients from one tier per round, alternating tiers over time. Although the original TiFL evaluation includes heterogeneity through differences in CPU compute capacity, it does not model GPU heterogeneity and is not representative of real-world device populations. We use the \emph{uniform} selection policy as detailed in the original work.

\begin{table}[h!]
\centering
\caption{Federated learning results on CIFAR-10 comparing Oort and TiFL selection strategies across four delay distributions. \textit{Max Time} is the maximum client compute time. \textit{Avg Time} is the mean compute time of selected clients with $\pm$ standard deviation. \textit{Total Time} is total FL wall-clock time. \textit{TT90 (Time to 90\%)} shows at which round convergence reaches 90\% of the best metric observed across all configurations. Spearman $\rho$ measures correlation between client speed and selection frequency: positive values indicate preference for faster clients.}
\begin{tabular}{l|ccccc}
\midrule
\textbf{Timing Source} & \textbf{Avg Round Time (s)} & \textbf{Total Time (s)} & \textbf{Speedup} & \textbf{$\rho$} & \textbf{TT90} \\
\midrule
\multicolumn{6}{l}{\textbf{OORT}} \\
  static       & 1.41 $\pm$ 0.18 & 14.15 & 1.000$\times$ & 0.202 $\pm$ 0.128 & 9 \\
  pareto       & 1.72 $\pm$ 1.51 & 17.16 & 0.824$\times$ & 0.341 $\pm$ 0.127 & 8 \\
  halfnorm     & 1.59 $\pm$ 0.42 & 15.94 & 0.887$\times$ & 0.228 $\pm$ 0.139 & 8 \\
  emulated     & 1.48 $\pm$ 0.78 & 14.82 & 0.954$\times$ & 0.324 $\pm$ 0.159 & 8 \\
\midrule
\multicolumn{6}{l}{\textbf{TIFL}} \\
  static       & 1.43 $\pm$ 0.31 & 14.25 & 1.000$\times$ & 0.000 $\pm$ 0.635 & 9 \\
  pareto       & 2.62 $\pm$ 2.34 & 26.17 & 0.545$\times$ & -0.036 $\pm$ 0.597 & 9 \\
  halfnorm     & 1.69 $\pm$ 0.77 & 16.86 & 0.845$\times$ & -0.005 $\pm$ 0.629 & 8 \\
  emulated     & 1.78 $\pm$ 1.06 & 17.76 & 0.802$\times$ & 0.031 $\pm$ 0.566 & 9 \\
\midrule
\end{tabular}
\label{tab:selection-results-cifar10}
\end{table}

\begin{table}[h!]
\centering
\caption{Federated learning results on LLM finetuning (DistilGPT-2) comparing Oort and TiFL selection strategies across four delay distributions}
\begin{tabular}{l|ccccc}
\midrule
\textbf{Timing Source} & \textbf{Avg Round Time (s)} & \textbf{Total Time (s)} & \textbf{Speedup} & \textbf{$\rho$} & \textbf{TT90} \\
\midrule
\multicolumn{6}{l}{\textbf{OORT}} \\
  static       & 16.93 $\pm$ 1.41 & 203.12 & 1.000$\times$ & -0.036 $\pm$ 0.096 & 8 \\
  pareto       & 17.68 $\pm$ 14.00 & 212.16 & 0.957$\times$ & 0.183 $\pm$ 0.056 & 7 \\
  halfnorm     & 18.12 $\pm$ 5.17 & 217.48 & 0.934$\times$ & -0.072 $\pm$ 0.075 & 7 \\
  emulated     & 14.65 $\pm$ 6.62 & 175.84 & 1.155$\times$ & 0.185 $\pm$ 0.083 & 6 \\
\midrule
\multicolumn{6}{l}{\textbf{TIFL}} \\
  static       & 15.57 $\pm$ 6.07 & 186.81 & 1.000$\times$ & 0.000 $\pm$ 0.578 & 9 \\
  pareto       & 22.22 $\pm$ 20.30 & 266.69 & 0.700$\times$ & 0.000 $\pm$ 0.569 & 8 \\
  halfnorm     & 17.60 $\pm$ 10.83 & 211.17 & 0.885$\times$ & -0.000 $\pm$ 0.578 & 7 \\
  emulated     & 16.41 $\pm$ 8.90 & 196.91 & 0.949$\times$ & -0.000 $\pm$ 0.587 & 7 \\
\midrule
\end{tabular}
\label{tab:selection-results-llm}
\end{table}

\subsection{Ease of use and reproducibility}
To assess portability, we integrated the emulation layer into NVIDIA FLARE. This required only adapting deployment scripts and reconnecting communication interfaces between the FL framework and the emulation layer, without modifying any core emulation components. This indicates that the approach generalizes across FL frameworks with minimal engineering effort.

\end{document}